\documentclass[conference]{IEEEtran}
\IEEEoverridecommandlockouts
\usepackage{cite}
\usepackage{amsmath,amssymb,amsfonts}
\usepackage{algorithmic}
\usepackage{graphicx}
\usepackage{textcomp}
\usepackage{xcolor}

\newcommand{\ket}[1]{|#1\rangle}

\def\BibTeX{{\rm B\kern-.05em{\sc i\kern-.025em b}\kern-.08em
    T\kern-.1667em\lower.7ex\hbox{E}\kern-.125emX}}
\IEEEoverridecommandlockouts\IEEEpubid{\makebox[\columnwidth]{979-8-3195-0662-7/26/\$31.00 $\copyright$2026 IEEE \hfill}\hspace{\columnsep}\makebox[\columnwidth]{ }}
\begin{document}

\title{Transformer-Based Time-Series Inference of Lindblad Dynamics in Open Quantum Systems }

\author{\IEEEauthorblockN{Julian Guam, Jianqing Liu}
\IEEEauthorblockA{Department of Computer Science, North Carolina State University, Raleigh, NC \\
jli242@ncsu.edu, jliu96@ncsu.edu}

\thanks{This work is supported in part by the National Science Foundation under grants 2304118 and 2326746.}
}

\maketitle

\begin{abstract}
The Lindblad master equation is the standard framework for describing the non-unitary evolution of open quantum systems, where environmental interactions induce dissipation and decoherence. When both the system Hamiltonian and the dissipation rates are partially unknown or explicitly time-dependent, traditional analytical inversion and system-identification techniques become intractable. Recent works have demonstrated that Transformer-based models can infer unknown dissipation rates from observable time series, yet these approaches typically rely on hand-crafted statistical features under idealized and highly restricted conditions. Here we advance the paradigm by introducing a raw time-series Transformer that directly ingests the full trajectories of Pauli expectation values $\langle\sigma _x(t)\rangle$, $\langle\sigma _y(t)\rangle$, and $\langle\sigma_z(t)\rangle$, thereby fully exploiting the self-attention mechanism for temporal modeling. The architecture is further extended to jointly learn unknown Hamiltonian parameters, handle multiple dissipation channels, and operate robustly under realistic measurement noise. Across all tested scenarios the model achieves consistently high reconstruction accuracy while eliminating manual feature engineering. This provides a scalable, robust, and versatile framework for quantum environment sensing in realistic open quantum systems. 
\end{abstract}


\section{Introduction}

The rapid development of quantum technologies has brought the field to a critical juncture~\cite{liu2025road,le2026non}. A central obstacle is the presence of noise and dissipation arising from unavoidable interactions between the quantum system and its environment. 
In practice, the dynamics of open quantum systems are commonly described by the Lindblad master equation, which combines coherent evolution generated by a Hamiltonian with dissipative processes characterized by jump operators and their associated rates. In many experimental settings, however, these dissipation rates are not known precisely in advance~\cite{le2025optimized}. They can vary over time due to drifting control electronics, temperature fluctuations, 1/f noise, or other environmental factors. Accurate knowledge of these time-dependent rates is essential for several reasons: it enables the design of effective error mitigation and dynamical decoupling protocols, supports the development of feedback-based quantum control~\cite{guo2025two,guo2026quantum}, and allows researchers to diagnose and improve device performance. As quantum systems grow in size and complexity, manual or semi-manual calibration procedures become increasingly time-consuming and impractical, creating a pressing need for automated, data-driven methods that can extract dissipation parameters directly from experimental observations.

Traditional approaches to this problem, such as quantum process tomography and optimization-based fitting of Lindblad models~\cite{samach2022lindblad}, face significant limitations. Full process tomography scales exponentially with the number of qubits and is highly sensitive to noise, while optimization-based methods often require strong assumptions about the form of the dynamics or the availability of high-quality initial states. Recent work has begun to explore machine learning as a scalable alternative. In particular, Chen et al. showed that Transformer architectures can infer time-dependent dissipation rates from observable time series with high accuracy~\cite{chen2025unraveling}. However, their approach relied on hand-crafted statistical and spectral features, assumed a known system Hamiltonian, and did not explicitly model the combined effects of state-preparation and measurement noise that are ubiquitous in real experiments.

In this report, we present a Transformer-based framework that addresses these limitations by operating directly on raw time-series trajectories of observables. Our approach eliminates the need for manual feature engineering, enables joint inference of unknown Hamiltonian parameters together with time-dependent dissipation rates, and incorporates realistic models of both preparation and measurement noise during training. We demonstrate that the resulting model achieves state-of-the-art reconstruction accuracy on standard single-qubit benchmarks, successfully recovers both dissipation coefficients and unknown coherent parameters in a joint inference setting, and maintains high performance under combined noise sources when provided with moderately larger training datasets. We further show that the framework can accommodate different parameterizations of the time-dependent rate (including both Bernstein polynomials and sinusoids) and that simple reparameterization can resolve intrinsic ambiguities that arise in periodic models.

The remainder of this paper is organized as follows. Section~\ref{sec:pre} provides background on system Hamiltonians, the Lindblad master equation, and the Transformer architecture. Section~\ref{sec:method} describes the proposed methodology, including the data generation pipeline with explicit noise models and the training/inference procedure. Section~\ref{sec:performance} presents evaluation results across five experiment setups of increasing practicality yet complexity. Section~\ref{sec:conclude} draws the conclusion of the work.

\section{Preliminary}\label{sec:pre}

\subsection{Closed Quantum Systems and Hamiltonian Evolution}
When a quantum system is perfectly isolated from its environment, its evolution is unitary and governed by the Hamiltonian operator H. The Hamiltonian encodes the energy of the system and the interactions among its components. For a closed system the state evolves according to the Schrödinger equation (for pure states) or the von Neumann equation (for density matrices $\rho$):
$$i\frac{d\rho}{dt}=[H,\rho]$$
(or equivalently $\rho(t) = U(t) \rho(0) U^\dagger(t)$ with $U(t) = e^{-iHt}$).

A very common simple Hamiltonian for a single qubit is $H = \sigma_z$, where $\sigma_z$ is one of the Pauli matrices: 
$$\sigma_z = \begin{pmatrix} 1 & 0 \\ 0 & -1 \end{pmatrix}.$$

This Hamiltonian describes a qubit with an energy splitting of 2 (in natural units where $\hbar = 1$) between the states $\ket{0}$ and $\ket{1}$. Physically, it corresponds to a spin-1/2 particle in a static magnetic field along the z-direction, or to a superconducting qubit biased at a particular frequency. 

Under this Hamiltonian the populations of $  |0\rangle  $ and $  |1\rangle  $ remain constant, but any coherence (superposition) between them acquires a relative phase that oscillates at frequency 2. On the Bloch sphere this appears as a steady precession of the Bloch vector around the z-axis. The direction of precession (clockwise or counterclockwise when viewed from above) is determined by the sign of the Hamiltonian: $  H = +\sigma_z  $ produces one sense of rotation, while $  H = -\sigma_z  $ produces the opposite sense. This precession is the quantum analog of Larmor precession familiar from classical magnetic resonance.

In the absence of any coupling to the environment, this evolution is completely deterministic and reversible. The state changes over time through the accumulation of relative phases, but no information is lost to the surroundings.

\subsection{Open Quantum Systems: The Reality of Noise and Dissipation}
No quantum system in the laboratory is perfectly isolated. Qubits interact with a vast number of environmental degrees of freedom—phonons in a solid, photons in the electromagnetic field, two-level fluctuators in materials, control electronics, and so on. These interactions cause decoherence: the loss of quantum coherence that is essential for quantum advantage. They also cause dissipation: the exchange of energy with the environment, leading to relaxation toward thermal equilibrium.
The full microscopic description of a qubit coupled to a large environment would require tracking an enormous number of environmental modes—an intractable task. Fortunately, under the Born-Markov approximation (weak coupling and memoryless environment), the reduced dynamics of the qubit alone can be described by a compact equation known as the Lindblad master equation.

\subsection{The Lindblad Master Equation}
The Lindblad (or Gorini–Kossakowski–Sudarshan–Lindblad) master equation is the most widely used framework for describing the dynamics of Markovian open quantum systems~\cite{manzano2020short}. For a single qubit it takes the form
$$\frac{d\rho}{dt} = -i [H, \rho] + \sum_k \gamma_k \left( L_k \rho L_k^\dagger - \frac{1}{2} \{ L_k^\dagger L_k, \rho \} \right)$$
where:
\begin{itemize}
    \item The first term is the coherent (unitary) evolution generated by the system Hamiltonian $  H  $.
    \item The summation describes dissipative processes. Each term is characterized by a jump operator $  L_k  $ and a non-negative rate $  \gamma_k \geq 0  $.
    \item The superoperator $  \mathcal{D}[L]\rho = L\rho L^\dagger - \frac12 \{L^\dagger L, \rho\}  $ is called the dissipator. It ensures that the evolution remains completely positive and trace-preserving, so that $  \rho(t)  $ stays a valid density matrix at all times.
\end{itemize}

Physically, the jump operators $  L_k  $ represent the ways in which the system can exchange quanta or information with the environment. Common examples for a qubit include:
\begin{itemize}
    \item $  L = \sigma_-  $ (lowering operator) with rate $  \gamma_-(t)  $: amplitude damping/energy relaxation (the qubit spontaneously emits energy into the environment and drops to a lower state).
    \item $  L = \sigma_z  $ with rate $  \gamma_\phi(t)  $: pure dephasing (loss of relative phase between $  |0\rangle  $ and $  |1\rangle  $ without energy exchange).
    \item Other operators can describe leakage to higher levels, correlated errors, etc.
\end{itemize}

When the rates $  \gamma_k  $ are constant, the equation describes time-homogeneous Markovian dynamics. In many realistic settings, however, the rates themselves can be time-dependent, $  \gamma_k(t)  $. This can arise from external control fields, fluctuating environments, or engineered time-varying couplings. The paper that forms the basis of the present work focuses precisely on the case of time-dependent dissipation rates.

Because the Lindblad equation is linear in $  \rho  $, it can be vectorized and solved numerically for small systems (a few qubits). For larger systems or when one wants to infer the unknown rates $  \gamma_k(t)  $ from experimental data, direct simulation becomes impractical, motivating machine-learning approaches.

\subsection{The Transformer Architecture}
Transformer is a neural network architecture introduced by Vaswani et al. in the landmark paper \textit{Attention is All You Need}~\cite{vaswani2017attention}. The original Transformer follows an encoder-decoder architecture. The encoder processes the input sequence and produces a rich contextual representation, while the decoder generates the output sequence (e.g., a translated sentence) by attending to both the encoder’s output and the previously generated tokens. Each encoder and decoder layer contains two main sub-layers: a multi-head self-attention mechanism and a position-wise feed-forward network. Residual connections and layer normalization are applied around each sub-layer to stabilize training.

In the context of this work, we utilize the encoder part of the Transformer architecture. Rather than generating sequences, our model performs regression: it takes as input a raw time series of observable expectation values and outputs a vector of coefficients that parameterize the unknown time-dependent dissipation rate. The self-attention mechanism is particularly well-suited to this task because it can learn to focus on the temporal patterns most relevant to inferring the underlying dynamics, without requiring manually engineered features.

\subsection{Learning Unknown Open-System Dynamics}
In a typical quantum experiment one cannot directly observe the full density matrix $  \rho(t)  $. Instead, one measures expectation values of observables such as the Pauli operators $  \sigma_x, \sigma_y, \sigma_z  $. These expectation values are functions of time that encode information about both the coherent Hamiltonian evolution and the dissipative processes acting on the system.

The central scientific question addressed by this work is the inverse problem: given a prepared quantum state and its time series of observables, can we recover the unknown time-dependent dissipation rates $  \gamma_k(t)  $? This is a form of system identification or quantum process tomography for open quantum systems, but performed in a data-driven, scalable way using modern machine learning methods.

Traditional approaches to this problem often rely on hand-crafted features extracted from the time series or on explicit fitting of a parameterized Lindblad model. Instead, the approach explored in this work replaces the feature engineering with Transformer that directly ingests raw time-series trajectories. By training on large ensembles of simulated trajectories generated from known Lindblad models (with randomized initial states and randomized dissipation parameters), the network learns to map observable dynamics back to the underlying dissipation rates.

This learning framework must contend with several realistic complications:
\begin{itemize}
    \item The initial state of the qubit is often unknown or only partially known.
    \item Experimental measurements contain noise (finite shot noise, readout errors, etc.).
    \item The functional form of $  \gamma_k(t)  $ may be unknown a priori; a flexible parameterization (such as Bernstein polynomials) is therefore useful.
    \item Multiple dissipation channels may act simultaneously.
\end{itemize}

The subsequent sections of this report describe the concrete implementation of such a Transformer-based learner, the data-generation pipeline using the QuTiP quantum simulation library~\cite{lambert2026qutip}, architectural choices, training procedures, and extensive numerical experiments that quantify how well the model recovers dissipation rates under varying conditions of systems, noise, data volume, and model complexity.

\section{Methodology}\label{sec:method}
This section describes the overall methodology developed to infer time-dependent dissipation rates in open quantum systems from observable trajectories using a Transformer-based neural network. The approach builds upon the framework introduced by Chen et al. in~\cite{chen2025unraveling} while addressing several of its limitations, most notably the reliance on hand-crafted input features, the assumption of a known system Hamiltonian, and the limited explicit treatment of realistic noise sources. The methodology comprises three main components: a data-generation pipeline that produces noisy observable trajectories from known Lindblad dynamics, a Transformer architecture that operates directly on raw time-series inputs, and a training procedure designed to learn a robust mapping from observed dynamics to the underlying dissipation parameters.

\subsection{Overview of the Approach}
The central task is formulated as a supervised regression problem. Given a time series of expectation values of the Pauli operators $  \langle\sigma_x(t)\rangle  $, $  \langle\sigma_y(t)\rangle  $, and $  \langle\sigma_z(t)\rangle  $ for a single qubit, the model is trained to predict the coefficients that parameterize the unknown time-dependent dissipation rate $  \gamma(t)  $. The system Hamiltonian is taken to be $  H = \sigma_z  $ (a conventional choice corresponding to a fixed energy splitting), while the dissipation is modeled via the standard amplitude-damping channel with a time-varying rate. To increase flexibility and physical realism, the rate $  \gamma(t)  $ is parameterized using a low-order Bernstein polynomial basis, which guarantees non-negativity over the finite time interval of interest while remaining expressive enough to capture a range of smooth temporal profiles.

The overall pipeline consists of three stages executed during training: (i) generation of synthetic trajectories by numerically integrating the Lindblad master equation with randomly sampled initial states and dissipation parameters, (ii) corruption of both the initial state and the resulting expectation values with realistic noise models, and (iii) supervised training of the Transformer to regress the dissipation coefficients from the noisy observable time series. At inference time, the trained model receives a new (noisy) trajectory and outputs the predicted coefficients without requiring knowledge of the initial state or explicit feature engineering.

\subsection{Data Generation Pipeline}
Synthetic training and validation data are generated using the QuTiP library for numerical integration of the Lindblad master equation. For each training example, either Bernstein polynomials of degree of two or sinusoidal functions are used as the time-dependent dissipation rate. For Bernstein polynomials, a three-dimensional coefficient vector $  (a_0, a_1, a_2)  $ is first sampled uniformly from a physically plausible interval that ensures $  \gamma(t) \geq 0  $ throughout the simulation window $  [0, T]  $. The time-dependent rate is then constructed as the quadratic Bernstein polynomial
$$\gamma(t) = a_0 b_{0,2}(t/T) + a_1 b_{1,2}(t/T) + a_2 b_{2,2}(t/T)$$
where $  b_{j,2}(s)  $ denotes the Bernstein basis polynomials of degree two. For sinusoidal functions, the coefficient vector consists of amplitude $(a_0)$, frequency $(f)$, and phase $(\phi)$ is sampled to construct 
$$\gamma(t) = a_0 \frac{1 + \sin(2\pi f t + \phi)}{2}$$
while ensuring $  \gamma(t) \geq 0  $. Both parameterizations provide a compact yet flexible representation of smooth, non-negative rate functions and has been shown to be effective for both constant and time-varying dissipation profiles. 

An initial pure state $  \rho_0  $ is drawn from the Haar measure on the single-qubit Hilbert space, corresponding to a uniformly random pure state on the Bloch sphere. The initial state is then evolved under the Lindblad dynamics
$$\dot{\rho}(t) = -i[H, \rho(t)] + \gamma_-(t)\mathcal{D}[\sigma_-]\rho(t)$$
for a fixed duration $  T  $, where $  \mathcal{D}[L]\rho = L\rho L^\dagger - \frac12\{L^\dagger L, \rho\}  $ is the standard dissipator. Expectation values of the three Pauli operators are recorded at $  N_t  $ uniformly spaced time points, yielding a clean trajectory matrix of shape $  (N_t, 3)  $.

Measurement noise is subsequently added to each expectation value to simulate finite-shot statistics. The noise is modeled as zero-mean Gaussian with a standard deviation proportional to the absolute value of the underlying signal, thereby reflecting the heteroscedastic nature of shot noise. The resulting noisy trajectories, together with the ground-truth coefficient vectors, constitute the supervised training examples. Because the initial state is never provided to the model, the learned mapping must be robust to both preparation noise and the inherent variability of Haar-random initial conditions.

\subsection{Transformer-based Learning System}
The core of the learning system is an encoder-only Transformer that directly ingests the raw time-series matrix of noisy expectation values. This design choice deliberately avoids the hand-crafted feature extraction step employed in prior work, allowing the self-attention mechanism to discover relevant temporal patterns and long-range dependencies autonomously.

Each input sequence is first linearly projected into a higher-dimensional embedding space. Because the raw time series already encodes temporal order, learnable positional encodings are added to preserve sequence information. The embedded sequence is then processed by a stack of Transformer encoder layers, each comprising multi-head self-attention followed by a position-wise feed-forward network with residual connections and layer normalization. After the final encoder layer, a global average pooling operation collapses the sequence dimension, producing a fixed-length representation that is passed to a lightweight multilayer perceptron (MLP) regression head. The MLP outputs a three-dimensional vector corresponding to the predicted Bernstein coefficients $  (a_0, a_1, a_2)  $ or sinusoidal parameters $(a_0, f, \phi)$.

The architecture is deliberately kept relatively lightweight compared with large language models, reflecting the modest input dimensionality (three observable channels) and the structured nature of the regression task. 

\subsection{Models and Training Procedure}
Training is performed by minimizing the mean-squared error (MSE) between the predicted and ground-truth coefficient vectors after both have been standardized to zero mean and unit variance. Standardization is applied independently to each coefficient using statistics computed on the training set; the same transformation is used at evaluation time. This normalization improves training stability and ensures that coefficients of different scales contribute comparably to the loss. Optimization is carried out with the AdamW optimizer together with a reduce learning-rate on plateau scheduler. Early stopping based on validation MSE is employed to prevent overfitting, with a patience window chosen to balance training time against generalization performance. 

The two noise models are parameterized by scalar hyperparameters that control their strength. Preparation noise is realized by applying a random unitary rotation whose angle is drawn from a zero-mean Gaussian whose standard deviation is treated as a tunable noise level. Measurement noise is realized by adding zero-mean Gaussian perturbations whose standard deviation at each time point is proportional to the absolute value of the clean expectation value at that point; the proportionality constant again serves as a tunable noise level. These two parameters allow systematic investigation of robustness across a range of realistic experimental conditions without altering the underlying Lindblad dynamics.

All data-generation, model, and training components are implemented in Python using QuTiP for quantum simulation, PyTorch for the neural network, and scikit-learn for preprocessing and evaluation utilities. The entire pipeline is designed to be modular, enabling straightforward extension to multi-qubit systems, additional dissipation channels, or alternative parameterizations of the time-dependent rates.

\section{Performance Evaluation}\label{sec:performance}
We evaluate the proposed Transformer-based framework through a series of controlled experiments designed to assess both its baseline performance and its ability to handle more challenging and realistic scenarios. All models are trained on raw time-series trajectories of the Pauli expectation values without hand-crafted feature extraction. Unless otherwise stated, the system Hamiltonian is taken to be $  H = \sigma_z  $ and the dissipation is described by a single amplitude-damping channel $\sigma_-$ with time-dependent rate $  \gamma_-(t)  $. Performance is quantified by the coefficient of determination $  R^2  $ between the predicted and ground-truth parameters on held-out test sets.

\subsection{Baseline Performance on Single-Qubit Time-Dependent Dissipation}
We first recreate the single-qubit time-dependent dissipation setting using the Bernstein-polynomial parameterization of degree two for $  \gamma_-(t)  $. The model is trained on 5,000 trajectories generated from Haar-random initial states. In contrast to the original work in~\cite{chen2025unraveling}, which relied on a set of hand-crafted statistical and spectral features, this model receives the raw observable time series directly.

\begin{figure}[tp]
\centering
\includegraphics[width=\columnwidth]{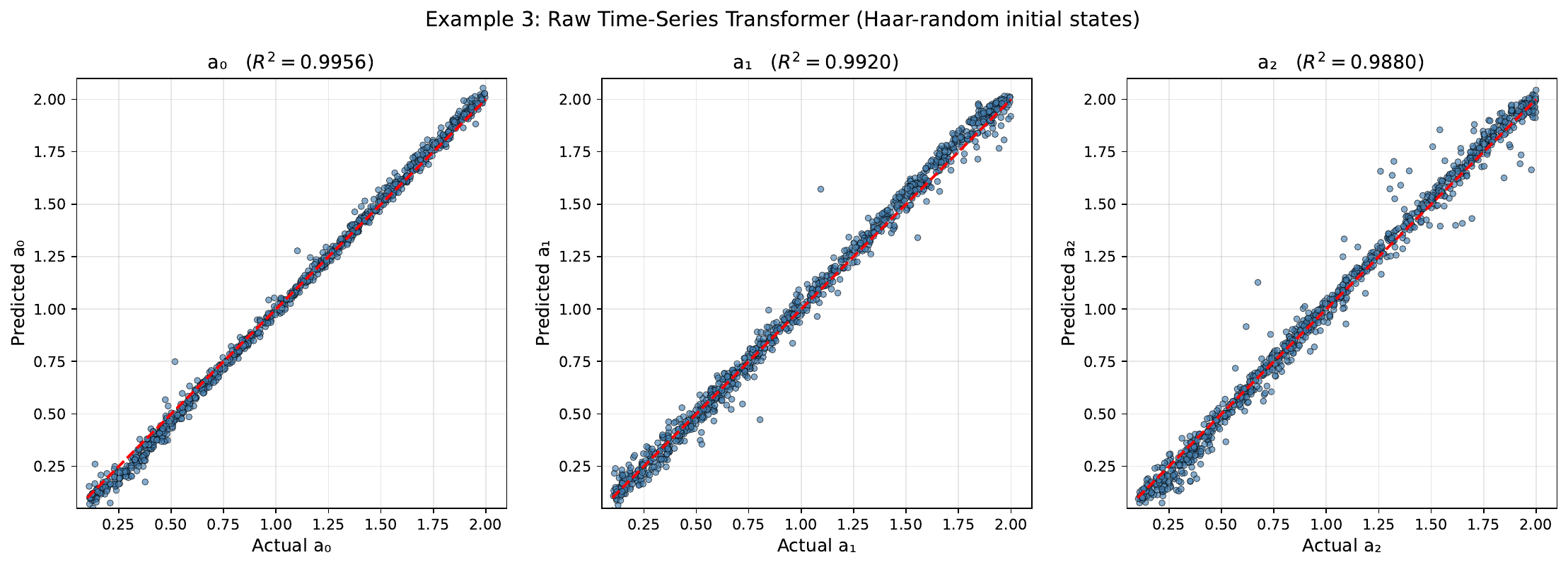}
\vspace{-0.2in}
\caption{Single-qubit time-dependent dissipation analysis}
\end{figure}

Under these conditions, the Transformer achieves $  R^2  $ scores of 0.9956, 0.9920, and 0.9880 for the three Bernstein coefficients $  a_0  $, $  a_1  $, and $  a_2  $, respectively. These values are consistently higher than those reported in the original study, demonstrating that the raw-sequence Transformer architecture can extract the necessary temporal information more effectively than the feature-engineered baseline.

\subsection{Joint Inference of an Unknown Hamiltonian Parameter and Dissipation Rates}
To test the framework’s ability to handle partially unknown coherent dynamics, we extend the Hamiltonian to $  H = \sigma_z + h \sigma_x  $, where the transverse-field strength $  h  $ is treated as an unknown scalar parameter to be inferred jointly with the dissipation coefficients. The model output dimension is increased accordingly to four, and training is performed on 5,000 trajectories with Haar-random initial states.

\begin{figure}[tp]
\centering
\includegraphics[width=\columnwidth]{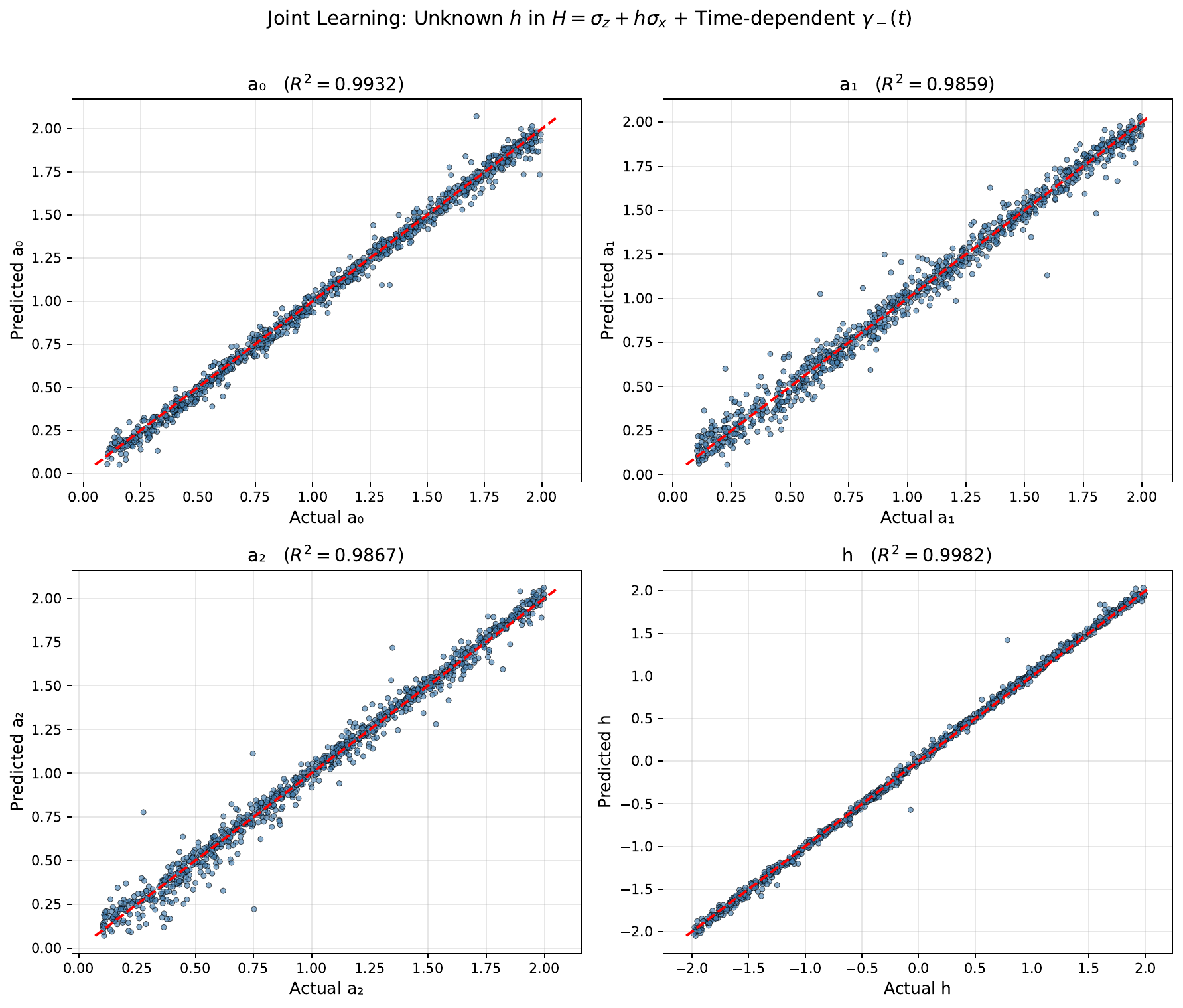}
\vspace{-0.2in}
\caption{Analysis of partially unknown Hamiltonian along with time-dependent dissipation}
\end{figure}

The resulting $  R^2  $ scores are 0.9932, 0.9859, and 0.9867 for the three dissipation coefficients, while the coefficient $  h  $ is recovered with an $  R^2  $ of 0.9982. The near-perfect recovery of the unknown Hamiltonian parameter indicates that the observable trajectories contain sufficient information to disentangle coherent and dissipative contributions, even when both are unknown a priori. The modest reduction in dissipation-coefficient accuracy relative to the known-Hamiltonian baseline remains within acceptable limits and confirms that the joint-inference task is feasible within the same architectural framework.

\subsection{Sinusoidal Parameterization of Time-Dependent Dissipation}
We next replace the Bernstein-polynomial parameterization with a sinusoidal form
$$\gamma_-(t) = a_0 \frac{1 + \sin(2\pi f t + \phi)}{2},$$
where the model is required to predict the parameters suitable for reconstructing the sinusoidal function. While the standard parameterization with the amplitude $  a_0  $, frequency $  f  $, and phase $  \phi  $ failed to produce any satisfactory results, alternative parameterizations prove to be feasible.
% With 5,000 training trajectories, the achieved $  R^2  $ values are 0.9776 for amplitude, 0.9669 for frequency, and 0.8231 for phase.
% \begin{figure}[tp]
% \centering
% \includegraphics[width=\columnwidth]{images/result3 sinusoidal.png}
% \vspace{-0.2in}
% \caption{Dissipation analysis using sinusoidal parameterization}
% \end{figure}

% The noticeably lower performance on the phase parameter is attributable to the inherent periodicity of the sine function. Because $  \sin(\theta + \phi) = \sin(\theta + \phi + 2\pi k)  $ for any integer $  k  $, many distinct phase values produce observationally indistinguishable trajectories over a finite time window. Consequently, the mapping from observable dynamics to phase is many-to-one, imposing a fundamental limit on the achievable $  R^2  $ that is independent of model capacity or data volume. This ambiguity is an intrinsic property of the chosen parameterization rather than a shortcoming of the learning architecture.

% \subsection{Improved Sinusoidal Parameterization Using the Initial Dissipation Rate}
We re-parameterize the sinusoidal rate by replacing the raw phase $  \phi  $ with the initial value $  \gamma_-(0)  $. The model now predicts the triple $  (a_0, f, \gamma_-(0))  $. With the same 5,000 trajectories, the $  R^2  $ scores achieve 0.9956 for amplitude, 0.9750 for frequency, and 0.9893 for $  \gamma_-(0)  $. The equally high performance compared to the baseline version with Bernstein-polynomials demonstrates that this transformer architecture is expressive and capable of reconstructing different rate functions without changing its model specifications, so long as appropriate parameterizations are chosen.
% The large gain in the third coefficient confirms that supplying an unambiguous, physically meaningful scalar at $  t = 0  $ removes the periodicity-induced degeneracy while preserving the expressive power of the sinusoidal model.

\begin{figure}[tp]
\centering
\includegraphics[width=\columnwidth]{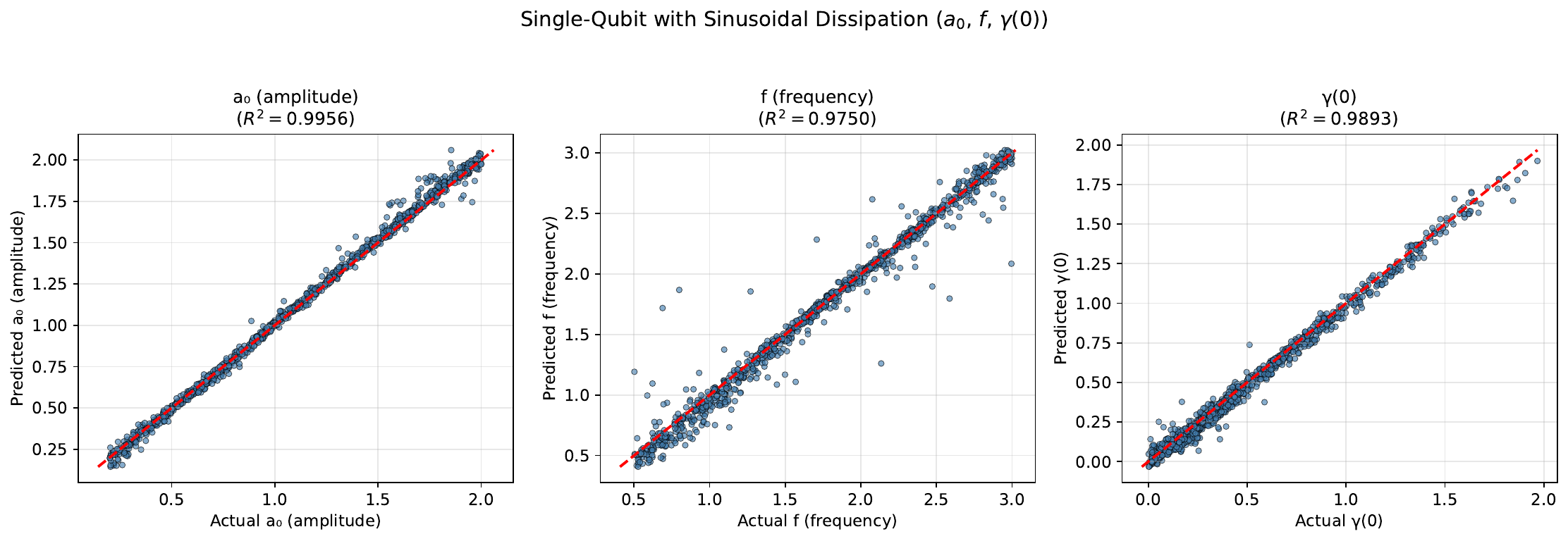}
\vspace{-0.2in}
\caption{Improved dissipation analysis by setting initial phases in sinusoidal parameterization}
\end{figure}

\subsection{Robustness to Measurement Noise}
We assess robustness to measurement noise by adding the expectation-value trajectories with additive Gaussian noise whose standard deviation is 2\% of the local signal amplitude, while keeping preparation noise at zero. Training is performed on 5,000 noisy trajectories using the original Bernstein parameterization. 
\begin{figure}[tp]
\centering
\includegraphics[width=\columnwidth]{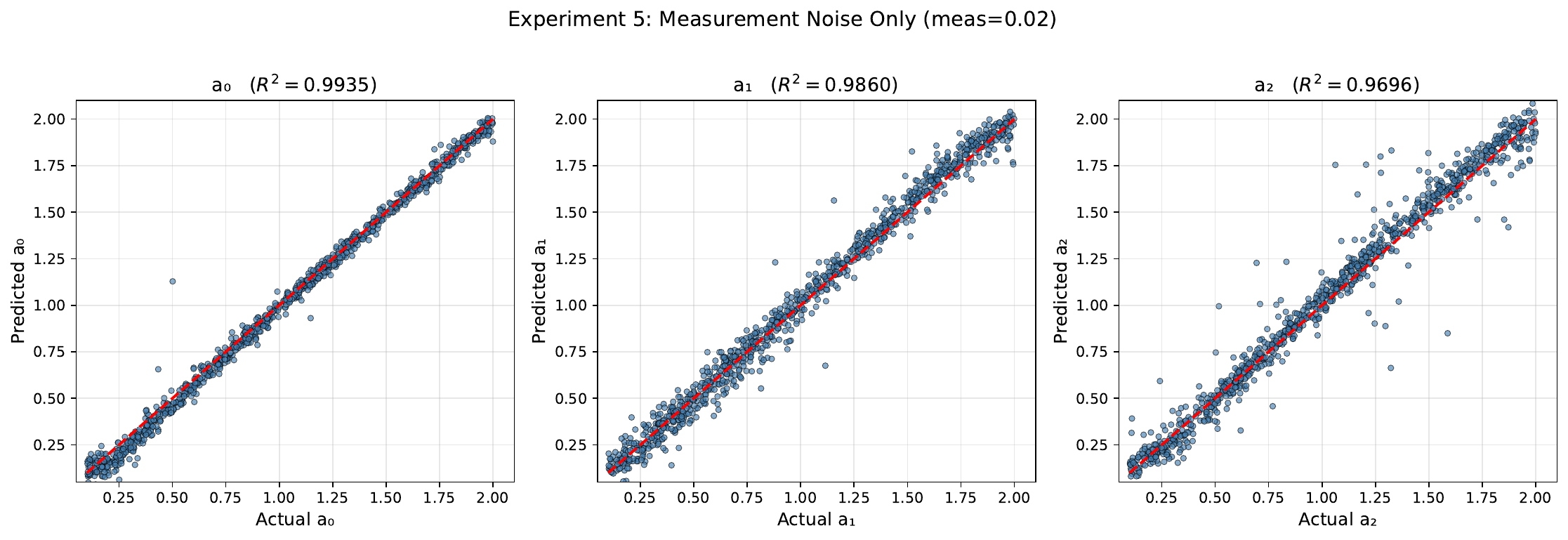}
\vspace{-0.2in}
\caption{Dissipation analysis with measurement noise}
\end{figure}

The resulting $  R^2  $ scores are 0.9935, 0.9860, and 0.9696 for $  a_0  $, $  a_1  $, and $  a_2  $, respectively — only marginally lower than the noise-free baseline. These results indicate that the Transformer can effectively average out moderate levels of shot noise across the time series without requiring architectural modifications or explicit denoising layers.

\subsection{Robustness to Preparation and Measurement Noise}
Finally, we evaluate performance under the joint action of preparation noise and measurement noise. A small random unitary perturbation (standard deviation 0.015 rad) is applied to the initial state, and measurement noise with relative amplitude 0.02 is added to the observable trajectories. To compensate for the increased stochasticity, the training set size is increased to 8,000 examples. 
\begin{figure}[tp]
\centering
\includegraphics[width=\columnwidth]{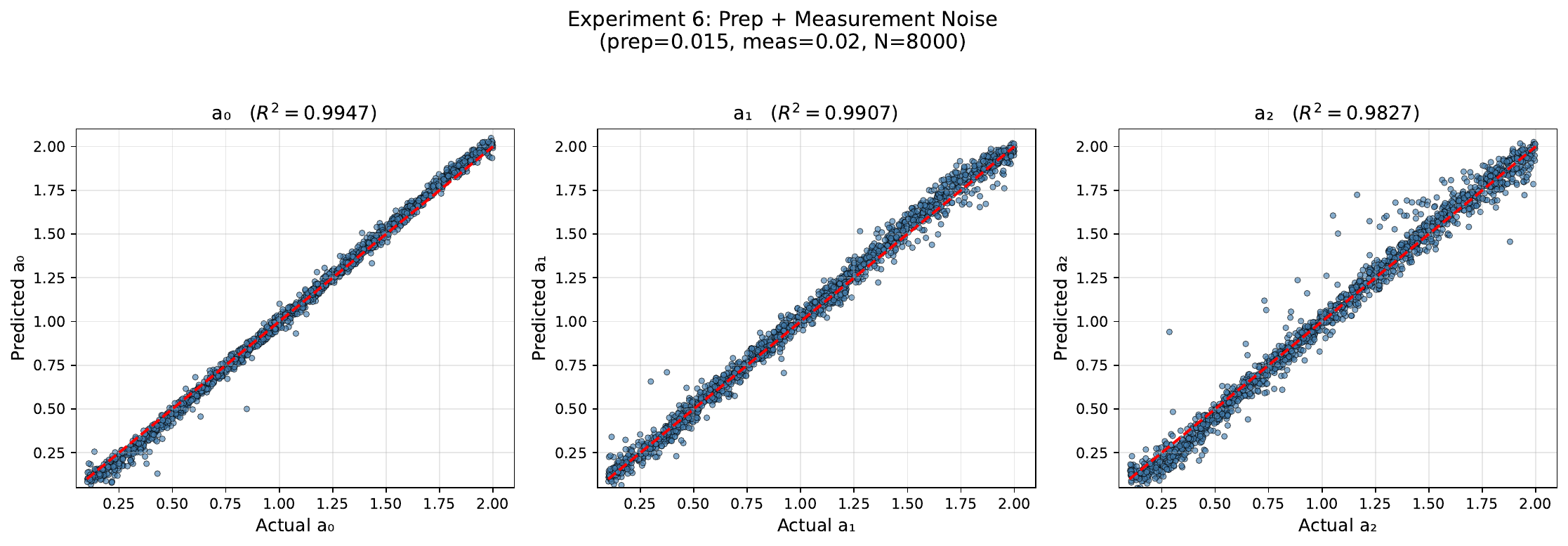}
\vspace{-0.2in}
\caption{Dissipation analysis with measurement and state preparation noise}
\end{figure}

Under these conditions the model achieves $  R^2  $ scores of 0.9947, 0.9907, and 0.9827 for the three dissipation coefficients. The retention of high accuracy, particularly for the most sensitive coefficient $  a_2  $, demonstrates that the framework remains effective even when both state-preparation errors and finite-shot measurement noise are present simultaneously, provided sufficient training data are available.

\subsection{Summary of Experimental Findings}
Across the five experiments, the raw-sequence Transformer consistently achieves high-fidelity recovery of dissipation parameters under a variety of physically motivated conditions. The architecture outperforms the hand-crafted-feature baseline of~\cite{chen2025unraveling} on the canonical single-qubit task, successfully extends to joint Hamiltonian-parameter inference, accommodates both polynomial and sinusoidal rate parameterizations (with appropriate reparameterization to resolve ambiguities), and maintains strong performance in the presence of realistic preparation and measurement noise when trained on moderately larger datasets. These results collectively establish that direct processing of noisy observable time series with attention-based models constitutes a viable and scalable route for data-driven characterization of open quantum system dynamics.

\section{Conclusion}\label{sec:conclude}
In summary, this work demonstrates that Transformer-based models operating on raw observable trajectories constitute a powerful and practical tool for inferring time-dependent dissipation in open quantum systems. By achieving high accuracy across a range of increasingly realistic conditions—including unknown Hamiltonian parameters and combined preparation and measurement noise—the framework advances the state of data-driven quantum system identification and provides a solid foundation for future efforts aimed at characterizing and ultimately controlling complex open quantum devices.

\bibliographystyle{IEEEtran}
\bibliography{ref}

\end{document}